\documentclass{webofc}
\usepackage[varg]{txfonts}   

\usepackage{lineno}
\usepackage{graphicx}
\usepackage{dcolumn}
\usepackage{bm}
\usepackage[colorlinks=true,urlcolor=blue,citecolor=blue,linkcolor=blue]{hyperref}
\usepackage{amsmath}
\usepackage{amssymb}
\usepackage{amsfonts}
\usepackage{latexsym}
\usepackage[T1]{fontenc}
\usepackage[figuresright]{rotating}
\usepackage[version=3]{mhchem}

\def\kevc1{\ifmmode\mathrm{\ keV/{\mit c}}
          \else$\mathrm{\ keV/{\mit c}}$\fi}

\def\MeVc1{\,MeV/{\mit c}}

\def\mevc1{\ifmmode\mathrm{\ MeV/{\mit c}}
          \else$\mathrm{\ MeV/{\mit c}}$\fi}
\def\gevc1{\ifmmode\mathrm{\ GeV/{\mit c}}
          \else$\mathrm{\ GeV/{\mit c}}$\fi}
\def\GeVc1{\ifmmode\mathrm{\ GeV/{\mit c}}
          \else$\mathrm{\ GeV/{\mit c}}$\fi}
\def\kevc2{\ifmmode\mathrm{\ keV/{\mit c}^2}
          \else$\mathrm{\ keV/{\mit c}^2}$\fi}
\def\Mevc2{\ifmmode\mathrm{\ MeV/{\mit c}^2}
          \else$\mathrm{\ MeV/{\mit c}^2}$\fi}
\def\Gevc2{\ifmmode\mathrm{\ GeV/{\mit c}^2}
          \else$\mathrm{\ GeV/{\mit c}^2}$\fi}
\def\Gev2c2{\ifmmode\mathrm{\ GeV^2/{\mit c}^2}
          \else$\mathrm{\ GeV^2/{\mit c}^2}$\fi}
\def\Pgp{\ifmmode\math{p}
         \else$\mathrm{p}$\fi}
\def\Pagp{\ifmmode\mathrm{\overline{p}}
         \else$\mathrm{\overline{p}}$\fi}
\def\Pgn{\ifmmode\mathrm{n}
         \else$\mathrm{n}$\fi}
\def\Pagpn{\ifmmode\mathrm{\overline{n}}
         \else$\mathrm{\overline{n}}$\fi}
\def\Pp{\ifmmode\mathrm{p}
         \else$\mathrm{p}$\fi}
\def\Pap{\ifmmode\mathrm{\overline{p}}
         \else$\mathrm{\overline{p}}$\fi}

\def\Pn{\ifmmode\mathrm{n}
         \else$\mathrm{n}$\fi}
\def\Pan{\ifmmode\mathrm{\overline{n}}
         \else$\mathrm{\overline{p}}$\fi}
\def\Py{\ifmmode\mathrm{Y}
         \else$\mathrm{Y}$\fi}
\def\Pay{\ifmmode\mathrm{\overline{Y}}
         \else$\mathrm{\overline{Y}}$\fi}

\def\PgL{\ifmmode\mathrm{\Lambda }
          \else$\mathrm{\Lambda }$\fi}
\def\PagL{\ifmmode\mathrm{\overline{\Lambda }}
            \else$\mathrm{\overline{\Lambda }}$\fi}
\def\PgS{\ifmmode\mathrm{\Sigma }
          \else$\mathrm{\Sigma }$\fi}
\def\PagS{\ifmmode\mathrm{\overline{\Sigma }}
            \else$\mathrm{\overline{\Sigma }}$\fi}
\def\PgSp{\ifmmode\mathrm{\Sigma^+}
          \else$\mathrm{\Sigma^+}$\fi}
\def\PagSp{\ifmmode\mathrm{\overline{\Sigma^+}}
            \else$\mathrm{\overline{\Sigma^+}}$\fi}
\def\PgSm{\ifmmode\mathrm{\Sigma^-}
          \else$\mathrm{\Sigma^-}$\fi}
\def\PagSm{\ifmmode\mathrm{\overline{\Sigma^-}}
            \else$\mathrm{\overline{\Sigma^-}}$\fi}
\def\PgSn{\ifmmode\mathrm{{\Sigma }^0}
            \else$\mathrm{{\Sigma }^0}$\fi}
\def\PagSn{\ifmmode\mathrm{\overline{\Sigma }^0}
            \else$\mathrm{\overline{\Sigma }^0}$\fi}
\def\PgX{\ifmmode\mathrm{\Xi }
          \else$\mathrm{\Xi }$\fi}
\def\PagX{\ifmmode\mathrm{\overline{\Xi }}
            \else$\mathrm{\overline{\Xi }}$\fi}
\def\PgXm{\ifmmode\mathrm{\Xi^-}
          \else$\mathrm{\Xi^-}$\fi}
\def\PagXm{\ifmmode\mathrm{\overline{\Xi^-}}
            \else$\mathrm{\overline{\Xi^-}}$\fi}
\def\PagXp{\ifmmode\mathrm{\overline{\Xi }^+}
            \else$\mathrm{\overline{\Xi }^+}$\fi}

\def\PgOm{\ifmmode\mathrm{\Omega^-}
          \else$\mathrm{\Omega^-}$\fi}
\def\PagOm{\ifmmode\mathrm{\overline{\Omega^-}}
            \else$\mathrm{\overline{\Omega^-}}$\fi}
\def\PgOp{\ifmmode\mathrm{\Omega^+}
          \else$\mathrm{\Omega^+}$\fi}
\def\PagOp{\ifmmode\mathrm{\overline{\Omega }^+}
            \else$\mathrm{\overline{\Omega }^+}$\fi}

\def\PgLc{\ifmmode\mathrm{\Lambda_c}
          \else$\mathrm{\Lambda_c}$\fi}
\def\PagLc{\ifmmode\mathrm{\overline{\Lambda }_c}
            \else$\mathrm{\overline{\Lambda }_c}$\fi}

\def\PgD{\ifmmode\mathrm{D}
          \else$\mathrm{D}$\fi}
\def\PagD{\ifmmode\mathrm{\overline{D}}
            \else$\mathrm{\overline{D}}$\fi}

\def\PgPi{\ifmmode\mathrm{\pi }
          \else$\mathrm{\pi }$\fi}
\def\PagPi{\ifmmode\mathrm{\overline{\pi }}
            \else$\mathrm{\overline{\pi }}$\fi}

\begin{document}
\title{Exploring the neutron skin by hyperon--antihyperon production in antiproton--nucleus interactions}

\author{\firstname{Martin} \lastname{Christiansen}\inst{1} \and
        \firstname{Falk} \lastname{Schupp}\inst{1} \and
         \firstname{Patrick} \lastname{Achenbach}\inst{1,2} \and
        \firstname{Michael} \lastname{B\"olting}\inst{1} \and
		 \firstname{Josef} \lastname{Pochodzalla}\inst{1,2}\fnsep\thanks{\email{pochodza@uni-mainz.de}} \and
		 \firstname{Marcell} \lastname{Steinen}\inst{1}
}

\institute{Helmholtz Institute Mainz, GSI Helmholtzzentrum f\"ur Schwerionenforschung, Darmstadt, Johannes Gutenberg University, 55099 Mainz, Germany
\and
Institute for Nuclear Physics, Johannes Gutenberg University, 55099 Mainz, Germany
          }

\abstract{%
In this work we propose a new method to measure the evolution of the neutron skin thickness between 
different isotopes. We consider antiproton--nucleus interactions close to the production threshold of $\Lambda \overline{\Lambda }$ and $\Sigma^-\overline{\Lambda }$ pairs. At low energies, $\Lambda \overline{\Lambda }$ pairs are produced in $\overline{\text{p}} +\text{p}$ collisions, while $\Sigma^-\overline{\Lambda }$ pairs can only be produced in $\overline{\text{p}} +\text{n}$ interactions. Measuring these cross sections provides information on the neutron skin thickness.
}
\maketitle
\section{Introduction}
\label{sec:intro}

Protons and Neutrons are the building blocks of conventional atomic nuclei. Their distribution in 
nuclei are related to bulk properties of nuclear matter, which are often expressed in terms of a nuclear equation of state (EoS).  Indeed, the isospin dependence of the EoS correlates strongly with the distribution of neutrons in nuclei~\cite{doi:10.1063/PT.3.4247}.

The neutron skin thickness, ${\Delta }r_\text{np}$, is defined as the difference
between the root-mean-squared (rms) point radii of the neutron and proton density distributions. The focus
of their studies often
lies on neutron rich doubly magic nuclei such as $^{40}$Ca, $^{48}$Ca, and $^{208}$Pb. However, it is the evolution of the neutron skin and proton distributions along isotope chains that provide important information for our understanding of the nuclear structure over the whole nuclear chart.

In this work we propose a new method to measure the difference of the neutron skin thickness between two different isotopes of a given element. We consider antiproton--nucleus interactions close to the thresholds of $\Lambda \overline{\Lambda }$ and $\Sigma^-\overline{\Lambda }$ pair production. At low energies, $\Lambda \overline{\Lambda }$ pairs are produced in $\overline{\text{p}}$+p collisions, while $\Sigma^-\overline{\Lambda }$ pairs can only be produced in $\overline{\text{p}}$+n interactions.
Measuring the probabilities $p_{\Lambda\overline{\Lambda }}$ and $p_{\Sigma^-\overline{\Lambda }}$ for the two processes for a reference isotope (I) and a second isotope (II), allows to determine the double ratio
\begin{equation}
  DR = \frac{p^{II}_{\Sigma^-\overline{\Lambda }}\Big/p^{II}_{\Lambda \overline{\Lambda }} }{p^{I}_{\Sigma^-\overline{\Lambda }}\Big/p^{I}_{\Lambda \overline{\Lambda }}}.
\label{eq:dr}
\end{equation}
Within a simple geometrical picture we show that this ratio
is strongly related to the difference of the neutron skin thicknesses of the two considered isotopes (I) and (II).
In a second step, we explore this ratio with the Gie\ss en Boltzmann--Uehling--Uhlenbeck transport model for a xenon isotope chain.
\begin{figure}
  \centering
  \includegraphics[width=7.5cm]{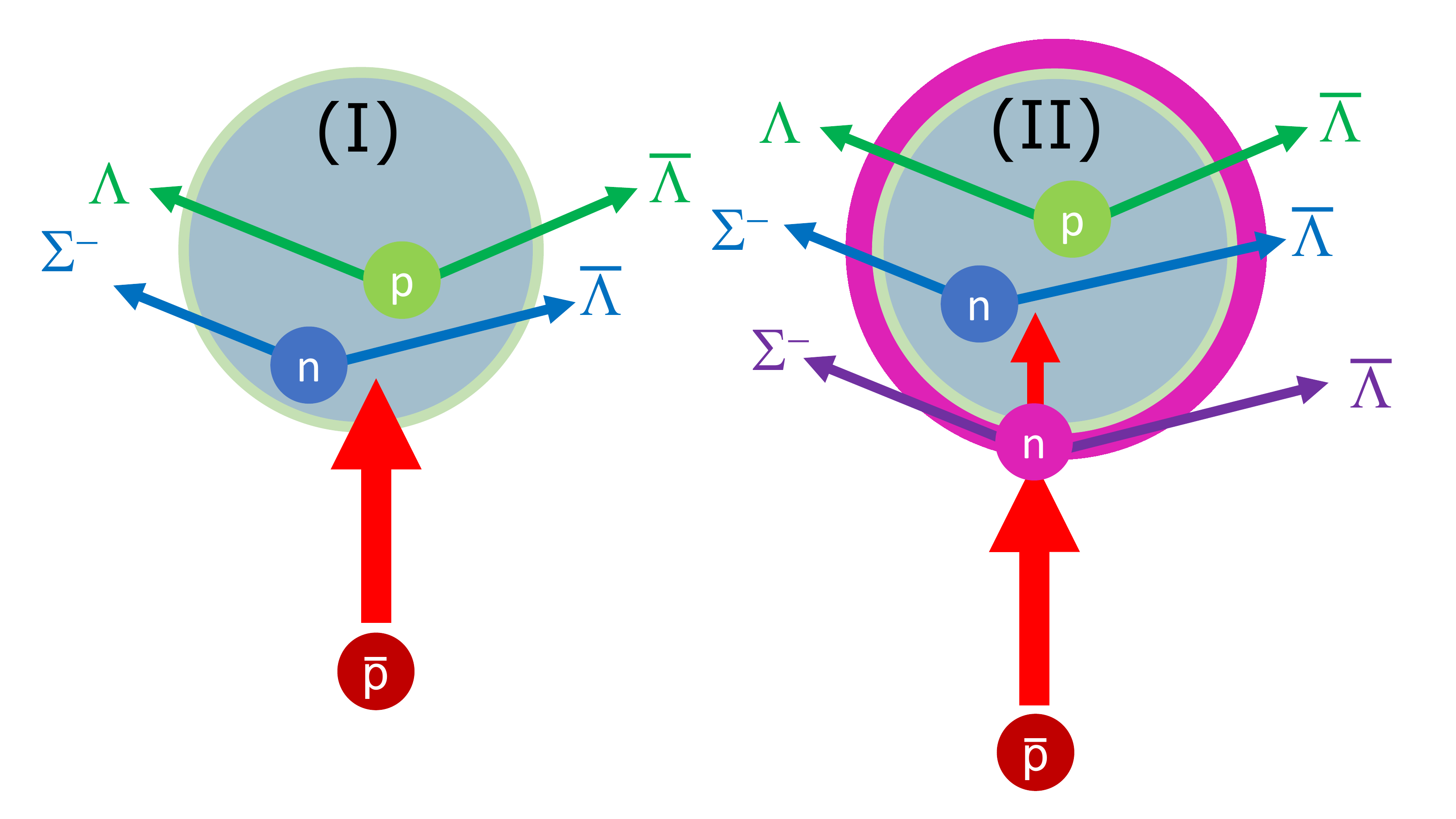}
  \caption{Illustration of $\Lambda \overline{\Lambda }$ and $\Sigma^-\overline{\Lambda }$ pair production in antiproton--nucleus interactions close to their thresholds.
  At low energies, $\Lambda \overline{\Lambda }$ pairs are produced in $\overline{\text{p}}$+p collisions, while $\Sigma^-\overline{\Lambda }$ pairs can only be produced in $\overline{\text{p}}$+n interactions.
   We consider two isotopes (I) and (II) which differ by an additional outer neutron layer with thickness $\Delta_\text{n}$, indicated in pink.
  }
  \label{fig:NeuS_scheme}
\end{figure}

\section{Motivation of the method}

Because of the strong absorption of antiprotons in nuclei the production of hyperon--antihyperon pairs happens in the nuclear periphery. For simplicity we consider a nearly central antiproton--nucleus collision (see left part of Fig.~\ref{fig:NeuS_scheme}). Close to threshold, the probability $p^{I}_{\Lambda \overline{\Lambda }}$ to produce a $\Lambda \overline{\Lambda }$ pair within the reference nucleus (I) can be written as
\begin{equation}
  p^{I}_{\Lambda \overline{\Lambda }}=\kappa_{\Lambda \overline{\Lambda }}        \cdot \frac{\rho (p)}{\rho (p)+\rho (n)}        \cdot \frac{\sigma_{\Lambda \overline{\Lambda }}}{\sigma_\text{tot}}.
\end{equation}
Here $\sigma_\text{tot}$  denotes the total ${\overline{\text{p}}+A}$ cross section
and $\sigma_{\Lambda \overline{\Lambda }}$ is the elementary $\overline{\text{p}}+{\text{p}}        \rightarrow \Lambda \overline{\Lambda }$ cross section.
$\rho (p)$ and $\rho (n)$ denote the effective densities in the periphery of the target nucleus of protons and neutrons, respectively. The factor $\kappa_{\Lambda \overline{\Lambda }}$ describes the loss of $\Lambda \overline{\Lambda }$ pairs by the absorption of the produced $\Lambda$ or $\overline{\Lambda }$ after their production.
Similarly, the production probability of $\Sigma^-\overline{\Lambda }$ pairs can be approximated by
\begin{equation}
  p^{I}_{\Sigma^-\overline{\Lambda }}=\kappa_{\Sigma^-\overline{\Lambda }}       \cdot \frac{\rho (n)}{\rho (p)+\rho (n)}                \cdot \frac{\sigma_{\Sigma^-\overline{\Lambda }}}{\sigma_\text{tot}}.
\label{eq:siglamI}
\end{equation}
Because of the large annihilation cross section of antibaryons in nuclei, both, $\kappa_{\Lambda \overline{\Lambda }}$ and $\kappa_{\Sigma^-\overline{\Lambda }}$ are dominated by the $\overline{\Lambda }$ absorption. Therefore, we assume
$\kappa_{\Lambda \overline{\Lambda }} \approx \kappa_{\Sigma^- \overline{\Lambda }} \equiv \kappa_I$.

We now turn to a isotope (II) with a larger neutron number and hence a more extended neutron distribution. For simplicity, we assume that the proton distribution remains identical to the one of nucleus (I) and that the neutron distribution is only extended by an additional halo $\Delta_\text{n}$ at the surface, see pink area in the right part of Fig.~\ref{fig:NeuS_scheme}. The production of $\Lambda \overline{\Lambda }$ pairs is reduced by the absorption probability $p_\text{abs}$ of the incident antiprotons within this additional neutron skin $\Delta_\text{n}$:
\begin{equation}
  p^{II}_{\Lambda \overline{\Lambda }}=(1-p_\text{abs})       \cdot \kappa_{II}       \cdot \frac{\rho (p)}{\rho (p)+\rho (n)} \cdot \frac{\sigma_{\Lambda \overline{\Lambda }}}{\sigma_\text{tot}}.
\label{eq:pll2}
\end{equation}
The absorption probability $p_\text{abs}$ can be expressed in terms of the total $\overline{\text{p}}+\text{n}$ reaction cross section
$\sigma_{\overline{\text{p}} \text{n}}$~\cite{PhysRevD.98.030001} and the integrated skin density $\int_{\Delta_\text{n}} \rho_\text{n}\, \text{d}r_\text{n}$ of the additional neutron skin of nucleus (II) with respect to the reference nucleus (I):
\begin{equation}
  1-p_\text{abs}\approx \exp \Big\{ -\sigma_{\overline{\text{p}}\text{n}}        \cdot \int_{\Delta_\text{n}} \rho_\text{n} \text{d}r_\text{n} \Big\}
\label{eq:pabs}
\end{equation}
Like the incoming antiprotons, the produced $\overline{\Lambda }$ are also absorbed in the additional neutron layer. We use the factorisation ansatz
$\kappa_{II}=\kappa_{I}\cdot \kappa_{n}$.
With this simplification we can express Eq.~(\ref{eq:pll2}) as
\begin{equation}
  p^{II}_{\Lambda \overline{\Lambda }} = \kappa_{n} \cdot (1-p_{abs}) \cdot p^{I}_{\Lambda \overline{\Lambda }}.
\label{eqLLII}
\end{equation}
The production of $\Sigma^-\overline{\Lambda }$ pairs gains an additional component from the
additional neutron skin. On the other hand, the contribution from the inner part of the nucleus (II)
is reduced by the loss of antiprotons in the additional neutron layer:
\begin{equation}
  p^{II}_{\Sigma^-\overline{\Lambda }} = \kappa_{II}                \cdot {p_\text{abs}} \cdot {\frac{\sigma_{\Sigma^-\overline{\Lambda }}}{\sigma_\text{tot}}} \\
+ (1-p_\text{abs}) \cdot \kappa_{II}\frac{\rho (n)}{\rho (p)+\rho (n)} \cdot \frac{\sigma_{\Sigma^-\overline{\Lambda }}}{\sigma_\text{tot}}.
\end{equation}
Here, we assume that the loss of outgoing $\Sigma^-$ and$/$or $\overline{\Lambda }$
for pairs produced in the additional neutron skin is the same as for pairs produced within the core nucleus. We thus obtain for the double ratio
\begin{equation}
  DR = \frac{p^{II}_{\Sigma^-\overline{\Lambda }}\Big/p^{II}_{\Lambda \overline{\Lambda }}}{p^{I}_{\Sigma^-\overline{\Lambda }}\Big/p^{I}_{\Lambda \overline{\Lambda }}}
  = \frac{p_\text{abs} \cdot \frac{\rho (p)+\rho (n)}{\rho (n)} + \Big(1-p_\text{abs}\Big) }{1-p_\text{abs}}
\end{equation}
With the assumption $\rho (p)$ = $\rho (n)$, we finally find for the double ratio the simple expression
\begin{equation}
  DR = \frac{1+p_\text{abs}}{1-p_\text{abs}}
\label{eq:dr01}
\end{equation}
To evaluate Eq.~(\ref{eq:dr01}), we use neutron skin thicknesses which are predicted during the initialization stage of the GiBUU simulations described in the next section
(cf.\ Fig.~\ref{fig:BUUradii}).
In the following we will compare the predictions of Eq.~(\ref{eq:dr01}) with the result of microscopic GiBUU simulations of $\overline{\text{p}} + A$ interactions at 2.4\GeVc1.

\section{Transport calculations for a Xe isotope chain}

\begin{figure}[bt]
\begin{minipage}[t]{0.48\textwidth}
  \includegraphics[width=\textwidth]{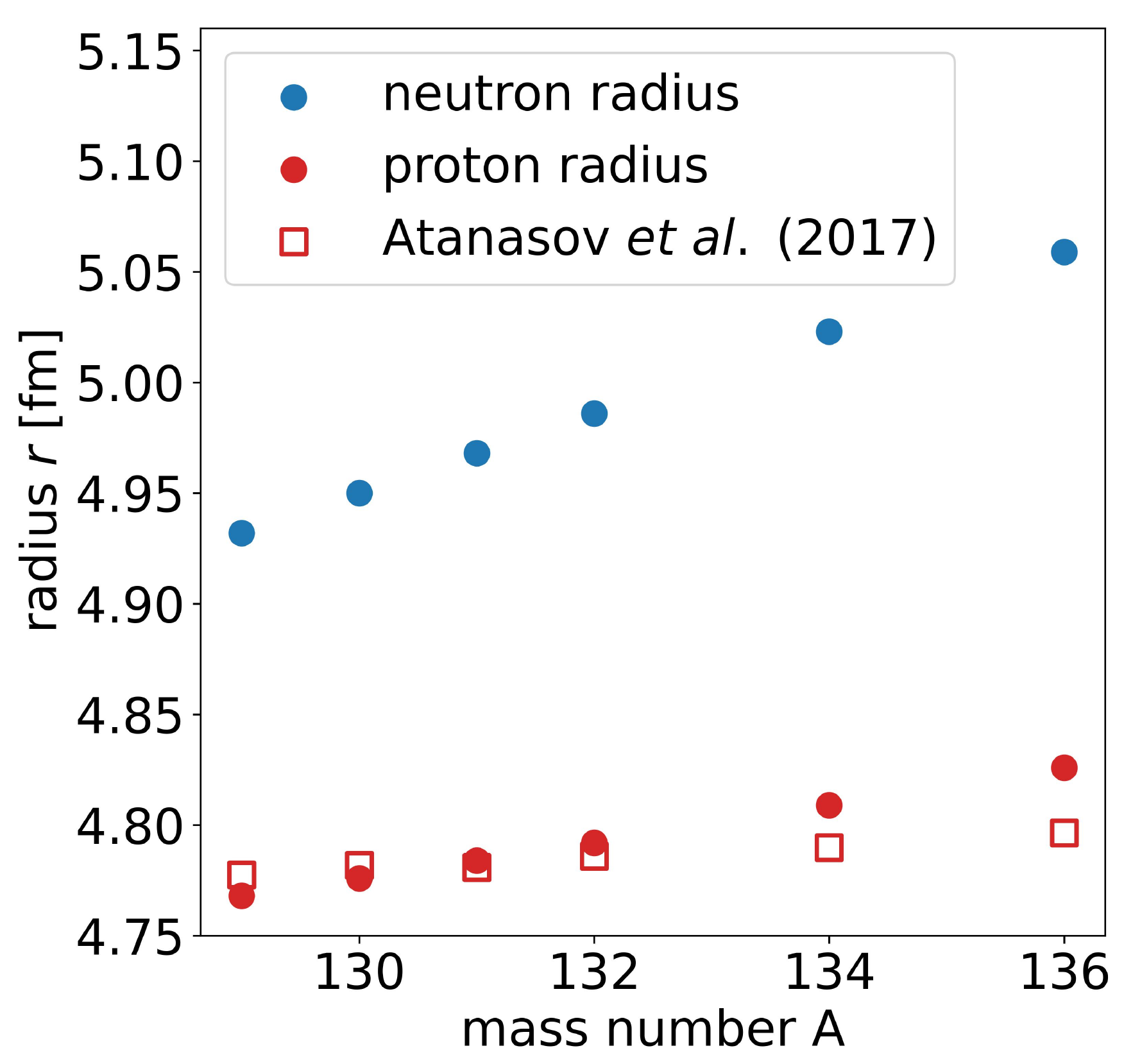}
  \caption{Proton (red) and neutron (blue) rms radii of the initial distributions for stable or long-lived xenon isotopes used by the GiBUU simulations. The red squares show experimental values for the proton radii~\cite{ANGELI201369,Atanasov_2017}. }
  \label{fig:BUUradii}
\end{minipage}
\hfill
\begin{minipage}[t]{0.48\textwidth}
  \includegraphics[width=\textwidth]{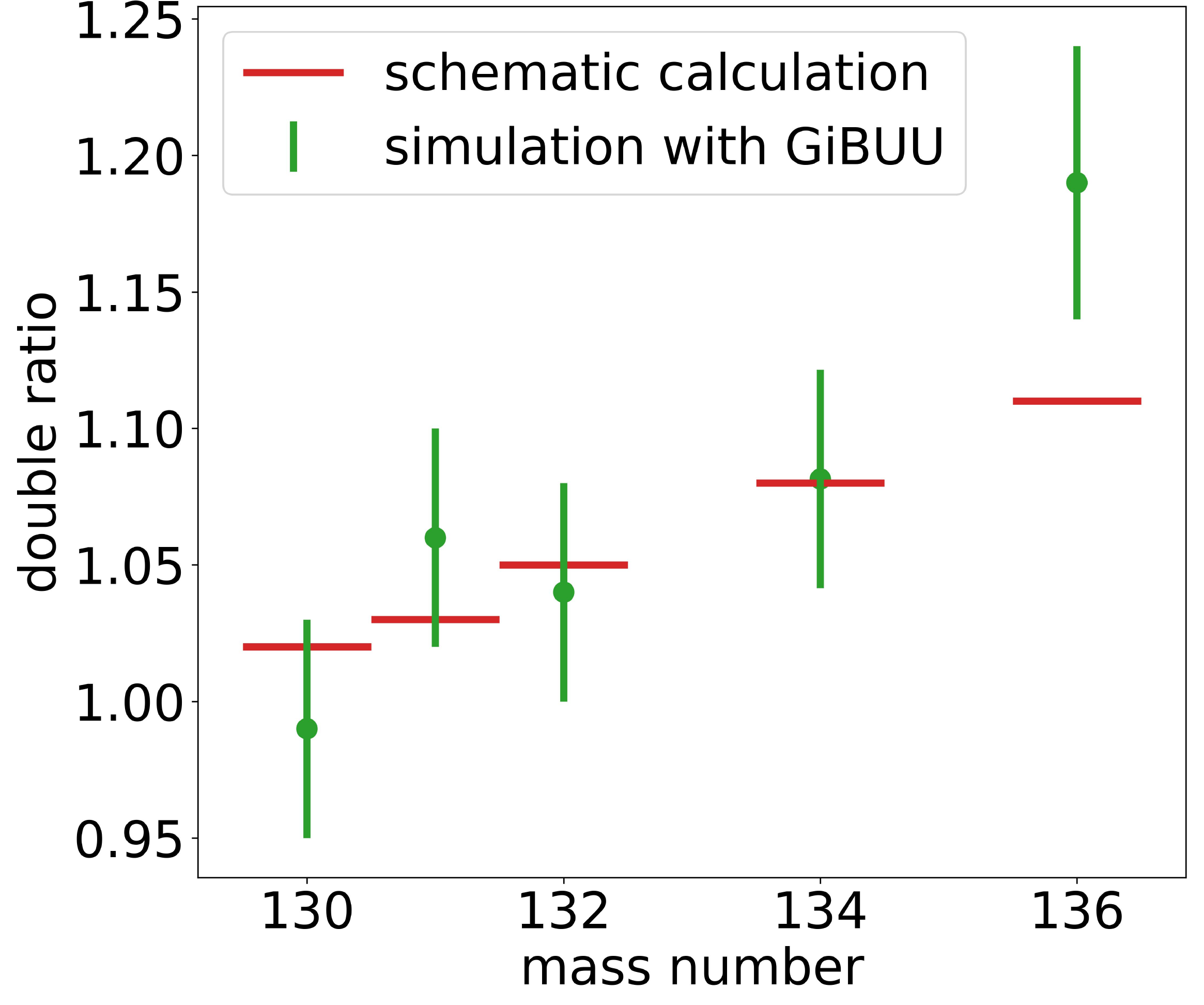}
  \caption{Green points show the double ratio predicted by the full GiBUU simulations. For comparison, the horizontal bars are the results of Eq.~(\ref{eq:dr01}), adopting the same proton and neutron radii as shown in Fig.~\ref{fig:BUUradii}.}
  \label{fig:BUUXe}
\end{minipage}
\end{figure}
Of course, this extremely simplified geometrical picture has several deficiencies. A more realistic description of the hyperon pair production can be achieved by microscopic transport calculations. As such a model, GiBUU describes many features of \Pagp--nucleus interactions in the FAIR energy range \cite{Buss20121} .

In the following we employ the GiBUU model to study $\overline{\text{p}}+\text{Xe}$ for a series of isotopes in the mass range from $A=129$ to $A=136$. These isotopes have a sizable abundance and can --- in principle --- be used for experimental studies. The simulations were performed with an incident antiproton momentum of 2.4 GeV/{\em{c}}. For each isotope about 28 million events were generated, requiring a computational time of about one day on the MOGON2 high performance computing cluster at the University of Mainz. During the initialization of these simulations, the proton and neutron distributions of the target nuclei were generated by a self consistent relativistic mean-field model.

The red and blue points in Fig.~\ref{fig:BUUradii} show the rms radii of the proton and neutron distributions as a function of the Xe mass number. In these calculations not only the neutron radius but --- unlike in the simple geometrical picture presented above --- also the proton radius is rising with increasing neutron number. This increasing charge radius is qualitatively consistent with experimental data, see open squares in Fig.~\ref{fig:BUUradii} \cite{ANGELI201369,Atanasov_2017}, though the experimental slope with the mass number is only about half as large.

The green points in Fig.~\ref{fig:BUUXe} show the double ratio DR predicted by the GiBUU simulations. For comparison, the horizontal bars in this figure show the results of Eq.~(\ref{eq:dr01}), using the same proton and neutron radii which are used as starting point for the GiBUU simulations. While the absolute agreement is less relevant,
we find a remarkable correlation between the double ratio evaluated with our simple analytical expression and the double ratio predicted by the complete GiBUU simulations. This correlation shows that the double ratio may be a good measure for the neutron skin difference. In future we will apply this method to the case of $^{40}$Ca and $^{48}$Ca.

\section*{Acknowledgements}
We thank Horst Lenske and Theo Gaitanos for valuable comments. This project has received funding from the European Union’s Horizon 2020 research and innovation programme under grant agreement No 824093. The presented data were in part collected within the framework of the bachelor theses of Martin Christiansen and the PhD thesis of Falk Schupp at the Johannes Gutenberg University Mainz.

\bibliography{lit_phys_hyp2022.bib}

\end{document}